\documentclass{article}

\usepackage{arxiv}

\usepackage[utf8]{inputenc} % allow utf-8 input
\usepackage[T1]{fontenc}    % use 8-bit T1 fonts
\usepackage{hyperref}       % hyperlinks
\usepackage{url}            % simple URL typesetting
\usepackage{booktabs}       % professional-quality tables
\usepackage{amsfonts}       % blackboard math symbols
\usepackage{nicefrac}       % compact symbols for 1/2, etc.
\usepackage{microtype}      % microtypography
\usepackage{lipsum}
\usepackage{graphicx}
\graphicspath{ {./images/} }
\usepackage{multirow}

\title{Towards "all-inclusive" Data Preparation to ensure Data Quality}

\author{
  Valerie Restat \\
  University of Hagen \\
  Hagen, Germany\\
  \texttt{valerie.restat@fernuni-hagen.de} \\
}

\begin{document}
\maketitle
\begin{abstract}
Data preparation, especially data cleaning, is very important to ensure data quality and to improve the output of automated decision systems. Since there is no single tool that covers all steps required, a combination of tools -- namely a data preparation pipeline -- is required. Such process comes with a number of challenges. We outline the challenges and describe the different tasks we want to analyze in our future research to address these. A test data generator which we implemented to constitute the basis for our future work will also be introduced in detail.
\end{abstract}

% keywords can be removed
\keywords{data preparation \and data quality \and data cleaning}

\section{Introduction}
\label{sec:intro}
The use of automated decision systems, e.g. based on machine learning, is increasing~\cite{Stoyanovich2020}. This impacts our everyday lives. Good quality data is the prerequisite for good quality decisions and predictions. To achieve this, data preparation, also referred to as data wrangling, data engineering or data preprocessing is necessary. This includes among other steps data profiling, data cleaning, data transformation~\cite{Klettke2021}, and data integration. As shown in ~\cite{Mahdavi2019} and~\cite{Li2021}, data cleaning is of particular importance for the improvement of machine learning based solutions.

Different tools exist for the individual preprocessing steps, which in turn support different algorithms. For example, KATARA~\cite{katara} can be used to detect pattern violations using knowledge bases while Dboost~\cite{dboost} integrates various algorithms for outlier detection. In addition, some tools cover multiple steps of data preparation. This includes proprietary tools such as Tableau Prep\footnote{\url{https://www.tableau.com/products/prep}} or Talend Data Preparation\footnote{\url{https://www.talend.com/products/data-preparation/}} as well as open source products like OpenRefine\footnote{\url{https://openrefine.org/}}.

However, no single tool covers all existing error types and data preparation tasks~\cite{Abedjan2016, Petrova_2020}. Different tools and algorithms need to be combined in a pipeline. This process poses a number of challenges.

The rest of the paper is structured as follows. In Section~\ref{sec:challenges}, we outline the current challenges regarding data preparation pipelines. A description of our research focus follows in Section~\ref{sec:research}. Different tasks that we will analyze in our future work will address those challenges. In Section~\ref{sec:gouda}, GouDa, our test data generator implemented as a basis for our analyses, will be introduced in greater detail. Section~\ref{sec:conclusion} summarizes the paper.

\section{Challenges of Data Preparation}
\label{sec:challenges}
A number of challenges arise during data preparation. These concern the input of the preparation, the preparation process itself and the result of this process. We published a first version of challenges and resulting requirements in~\cite{holistic-paper}.

\paragraph{Input of Data Preparation}
In the age of big data, more and more data is being stored and processed. In addition to the sheer \emph{volume} and \emph{velocity} of data, this also leads to a \emph{variety} of different data formats~\cite{Ridzuan2019}. 

\paragraph{Data Preparation Process}
The data preparation process or workflow poses many challenges. On the one hand, data preparation is mostly an \emph{iterative} process with an \emph{ad-hoc evaluation}~\cite{Krishnan2016}. On the other hand -- outlined in Section~\ref{sec:intro} -- \emph{no single dominant tool}, covering all existing error types and preprocessing tasks, exists yet. In addition, many tools require a preprocessed version of the data set, e.g. with a header and uniform delimiters, from the start~\cite{Hameed2020}. So far, there is a \emph{lack of intelligent solutions} and a great deal of \emph{human involvement} is still required~\cite{Hameed2020, Abedjan2016}. Hence, data preparation usually is a very \emph{time-consuming} process~\cite{Krishnan2016}.

Concerning big data, \emph{volume}, \emph{variety}, and \emph{velocity} generate further challenges~\cite{Ridzuan2019}. Many tools do not offer sufficient scalability. Moreover, varied data errors and data formats compromise data quality. The quality of the data is considered in greater detail below. Due to the velocity, streaming data and data changes must be taken into account in addition to static data.

Data preparation and new algorithms must be \emph{reproducible}~\cite{Pawlik2019}. In particular, the iterative nature of data preprocessing leads to further challenges in this context~\cite{Rupprecht2020}.

\paragraph{Result of Data Preparation}
The goal of data preparation is to ensure the best possible data quality. No single definition of data quality exists in literature. It is nevertheless undisputed that data quality addresses a wide range of different aspects. \emph{Accuracy} is one of the most referenced aspects. However, \emph{availability} and \emph{timeliness}, \emph{usability} as well as \emph{completeness} and \emph{relevance} are crucial factors too~\cite{Cai2015, Sidi2012}. Another important criterion for data quality is the evaluation in terms of \emph{bias and fairness}~\cite{Pitoura2020}. In context of fairness, \emph{explainability} is also a critical factor~\cite{Klettke2021_2}.

\begin{figure}[h]
  \centering
  \includegraphics[width=0.75\linewidth]{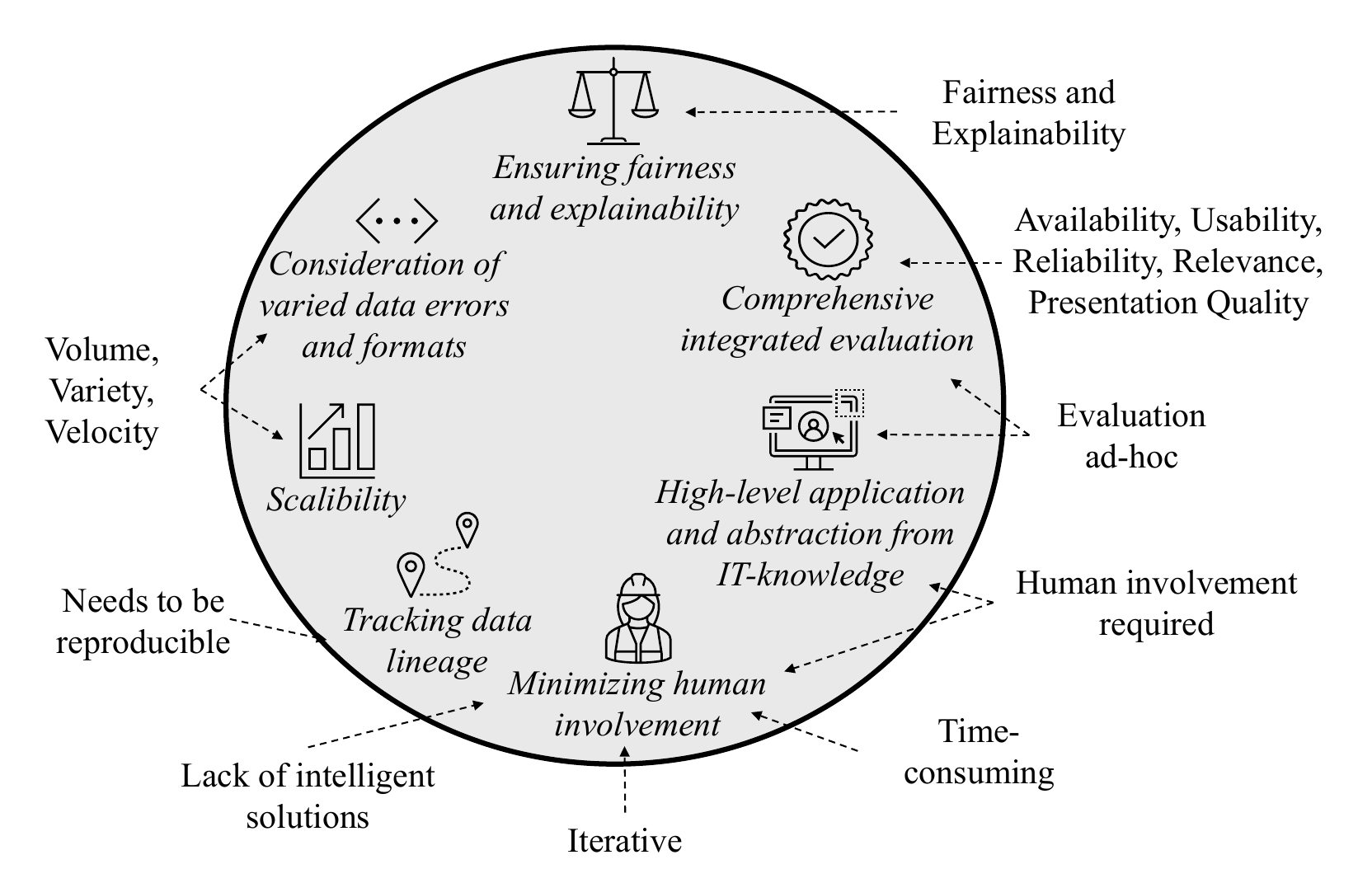}
  \caption{Holistic Data Preparation Tool}
  \label{fig:holistic_tool}
\end{figure}

\medskip
In order to meet the aforementioned challenges, we proposed a holistic data preparation tool~\cite{holistic-paper}, illustrated in~\autoref{fig:holistic_tool}. 
It reflects the challenges described and, obtained from them, the requirements for such a tool.  We will analyze these requirements in greater detail in Section~\ref{sec:research}. 

\section{Research Focus}
\label{sec:research}
\begin{figure}[h]
  \centering
  \includegraphics[width=0.7\linewidth]{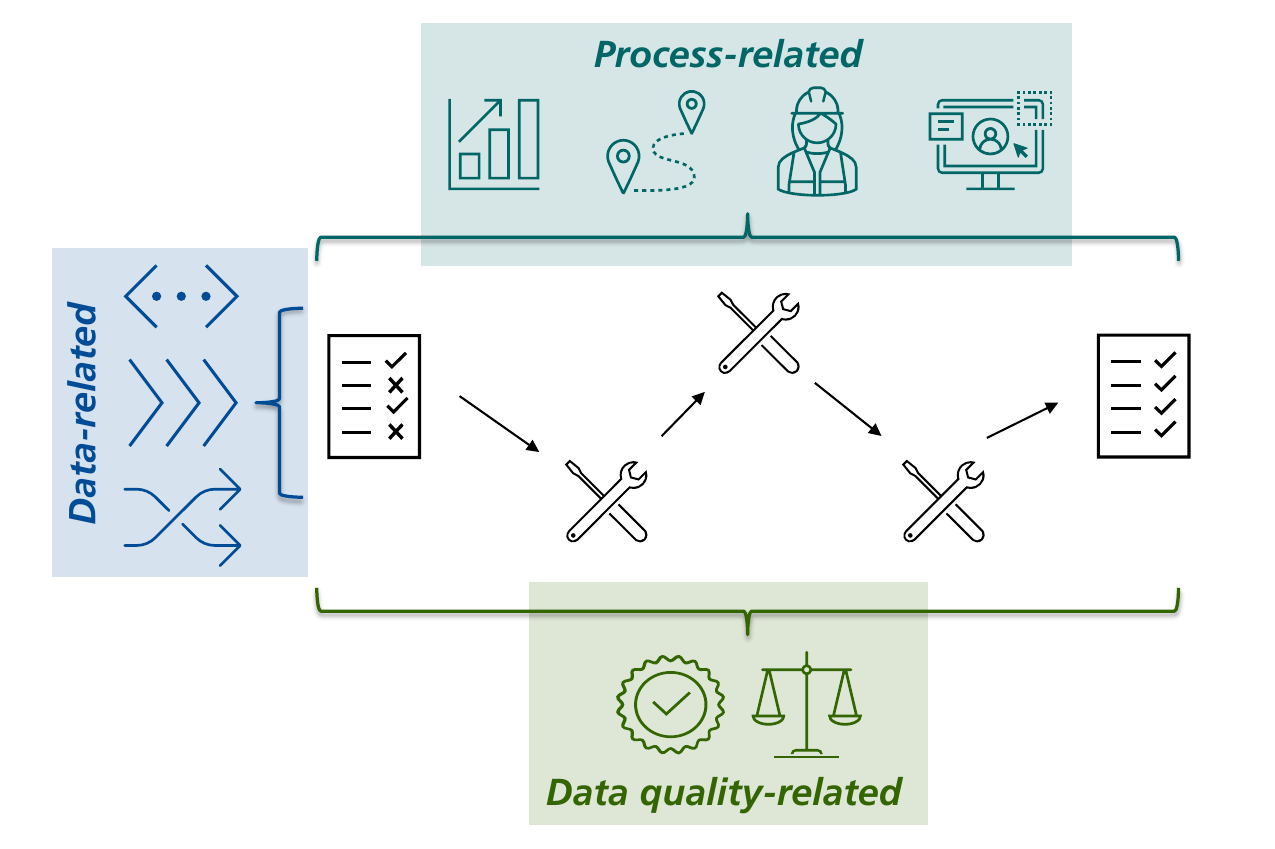}
  \caption{Research Focus}
  \label{fig:research_focus}
\end{figure}

In line with the challenges mentioned and the holistic data preparation tool presented, this section introduces specific research topics to be addressed in our future work. These can be divided into three parts, as illustrated in~\autoref{fig:research_focus}.

The first one is related to the goal of data preparation: data quality. Data quality-related aspects include ensuring fairness and a comprehensive evaluation. The second one focuses on how to achieve this goal: the process-related aspects. These include abstraction from IT-knowledge, minimizing of human involvement, and tracking data lineage to ensure reproducibility as well as scalability. The third one, on the other hand, concerns the data itself and their impact on both data quality and the process. Changes to the data and to streaming data as well as the formats of semi-structured and unstructured data are taken into account. The individual aspects are analyzed in detail below.

\subsection{Data quality-related}
At the outset, aspects related to data quality are considered. As described in Section~\ref{sec:challenges}, data quality includes a great variety of aspects and does not only deal with accuracy. Following topics were taken into consideration by us:

\bigskip
\paragraph{Ensuring fairness and explainability}
\begin{figure}[h!]
  \centering
  \includegraphics[width=0.1\linewidth]{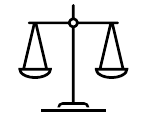}
  %\caption{Ensuring fairness}
  \label{fig:bias_fairness}
\end{figure}
Fairness is a crucial element in data quality. It must be ensured that there is no bias in the data. This particularly applies to automated decision systems~\cite{Stoyanovich2020}. In addition, explainability is also an important requirement~\cite{Klettke2021_2}.

Bias may already be present in the data, e.g. due to historical perceptions~\cite{Pitoura2020}. This can even be increased because automated decision systems such as machine learning models often reproduce trends and patterns in the data, reinforcing the bias~\cite{Klettke2021_2}. In addition, bias can also be introduced into the data by preprocessing itself, as shown in~\cite{Schelter2020}.

For a bias already present in the data, we want to analyze the applicability and interaction of different fairness metrics. Fairness is a very complex topic with many different types of bias in existence~\cite{Mehrabi2021}. We will investigate types of bias already covered by specific fairness metrics, identify unresolved issues and explore possible solutions.

For a bias arising during preprocessing, the detection shall be scrutinized in our future work. To combat bias, but also to ensure explainability, it is necessary to measure the extent to which preprocessing changes the data~\cite{Klettke2021_2}. Different avenues for the detection and measurement of changes in the data and distribution will be researched. Furthermore, we seek to identify those algorithms that generate the risk of introducing bias. This could make the use of consumer labels, compared to those envisioned in~\cite{Seifert2019}, conceivable.

\bigskip
\paragraph{Comprehensive integrated evaluation}
\begin{figure}[h!]
  \centering
  \includegraphics[width=0.095\linewidth]{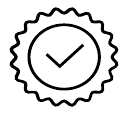}
  %\caption{Comprehensive integrated evaluation}
  \label{fig:evaluation}
\end{figure}

Since fairness is only one aspect of data quality, a comprehensive integrated evaluation is required. This should incorporate all the manifold aspects of data quality.

Accomplishing an automated, fully comprehensive evaluation adds to our scope of future research. To realize this, the impact of automated checks based on arrival time, range of data and accepted values will be examined. Different metrics, standards and specifications will be included in this project. The use of consumer labels is an additional consideration.

\subsection{Process-related}
Next, aspects concerning the preprocessing process itself shall be reviewed. 

\bigskip
\paragraph{High-level application and abstraction from IT-knowledge}
\begin{figure}[h!]
  \centering
  \includegraphics[width=0.1\linewidth]{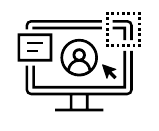}
  %\caption{High-level application and abstraction from IT-knowledge}
  \label{fig:abstraction_from_IT}
\end{figure}
As detailed in Section~\ref{sec:challenges}, human involvement is necessary for data preprocessing. Domain experts ought to be able to utilize tools without in-depth IT-knowledge which leads us to the conclusion that a significant demand for high-level applications exists.

To achieve abstraction from IT-knowledge, different approaches are conceivable, that nevertheless demand thorough analyses in our research. Visual and interactive approaches as well as approaches based on natural language processing and example-oriented concepts will be included.

\bigskip
\paragraph{Minimizing human involvement}
\begin{figure}[h!]
  \centering
  \includegraphics[width=0.075\linewidth]{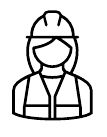}
  %\caption{Minimizing human involvement}
  \label{fig:human_involvement}
\end{figure}
Human involvement is still time-consuming and should be minimized. Thus, we will analyze how machine learning can be used to automatically preprocess data or generate suggestions for domain experts. The use of dictionaries and knowledge stores would also be conceivable for this purpose.

\bigskip
\paragraph{Data Lineage and Reproducibility}
\begin{figure}[h!]
  \centering
  \includegraphics[width=0.1\linewidth]{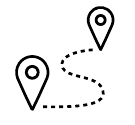}
  %\caption{Data Lineage and Reproducibility}
  \label{fig:reproducibility}
\end{figure}
Reproducibility is a very important aspect. The proposed tool should provide data lineage tracking capabilities, as well as rollback capabilities.

As shown in~\cite{Pawlik2019}, the raw data and already preprocessed data alone are insufficient for reproducibility. Our aim is to assess how the reproducibility of data cleaning pipelines can be guaranteed and to explore possibilities for the tracking of data lineage. Individual steps of the pipeline need to be made available to others efficiently so that they can be reproduced in the long term.

\bigskip
\paragraph{Scalibility}
\begin{figure}[h!]
  \centering
  \includegraphics[width=0.09\linewidth]{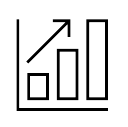}
  %\caption{Scalibility}
  \label{fig:scalibility}
\end{figure}
One of the preprocessing challenges posed by big data is volume. Handling the increasing amount of data requires scalable tools. 

We will analyze different possibilities to achieve efficient parallel execution and shall include a comparison of different parallel python data frames, such as Dask~\cite{Christ2016}, Modin~\cite{Petersohn2021}, and others.

\subsection{Data-related}
In this section, aspects regarding the data itself will be considered. There are three main aspects.

\bigskip
\paragraph{Support of semi-structured and unstructured data}
\begin{figure}[h!]
  \centering
  \includegraphics[width=0.1\linewidth]{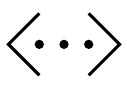}
  %\caption{Semi- and unstructured data}
  \label{fig:semi_unstructured_data}
\end{figure}
Most research in data preparation focuses on structured data~\cite{Chu2016}. Also, most tools are only suitable for structured data. The processing and data quality of semi-structured and unstructured data, on the other hand, has not yet been sufficiently researched.

This issue will be addressed in our research. On the one hand, we want to analyze the data quality of semi-structured and unstructured data. Many metrics like accuracy are only suitable for structured data. We therefore want to investigate which types of evaluation criteria and metrics are suitable for semi-structured and unstructured data.

On the other hand, we will investigate the processing of semi-structured and unstructured data. We want to analyze which methods and algorithms are suitable for which types of data and how a qualitative preprocessing of semi-structured and unstructured data can be ensured.

\bigskip
\paragraph{Data changes}
\begin{figure}[h!]
  \centering
  \includegraphics[width=0.1\linewidth]{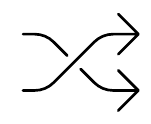}
  %\caption{Data changes}
  \label{fig:data_changes}
\end{figure}
Methods of data preparation are often based on statistical characteristics and rules. These may change over time which may cause problems. This is also referred to as \emph{concept drift}, a known problem especially in terms of machine learning models~\cite{Ditzler2013}. This can e.g. lead to the fact that integrity constraints may not hold~\cite{Volkovs2014}.

A future focus of our work will be on the analysis of how to handle changes in the data. This includes both semantic and syntactic changes. We want to investigate how to detect changes in the data and how to decide whether any preprocessing algorithms need to be adapted. Mean imputation is one example. If the mean value of an attribute changes significantly due to changes in the data, the imputation may need to be adjusted.

The different data formats also play a role here. In addition to structured data, semi-structured and unstructured data also require consideration.

\bigskip
\paragraph{Streaming data}
\begin{figure}[h!]
  \centering
  \includegraphics[width=0.1\linewidth]{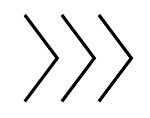}
  %\caption{Streaming data}
  \label{fig:streaming}
\end{figure}
More and more data come from sensors or mobile devices. Similar problems caused by data changes also occur with streaming data. Further difficulties arise since the total amount of data and its statistical characteristics are not known beforehand. New approaches to data preparation on distributed streams of data need to be explored in future~\cite{Chu2016}.

We want to analyze the difference between streaming data and static data sets. Accordingly, the identification of the special characteristics of streaming data compared to static data and their impact on data preparation contributes to our further research. We also want to examine which data cleaning approaches for static data sets are applicable to data streams. For scenarios where static data methods are not applicable, we want to explore new approaches.

\subsection{Related Work}
Other approaches also strive for holistic data preparation and analysis. Data Civilizer 2.0~\cite{Rezig2019}, for example, is a workflow system for data cleaning and parameter-tuning of machine learning models. Another example is KNIME~\cite{Fillbrunn2017}, an open source software that focuses on integration to help third-party developers embed their tools, regardless of the domain.

The proposed categories are helpful for comparing different tools and approaches. The goal of preprocessing, or the aspect of data quality targeted, varies depending on the use case. For example, fairness plays a very important role when it comes to personal data, but it is not relevant for sensor data. Data quality is thus independent of the the process-related aspects, but should always be taken into account. Data Civilizer 2.0 and KNIME do not provide information about data quality aspects.

The process-related aspects divide the tools depending on their functionality. For example, Data Civilizer 2.0 focuses on support for domain experts through visualization of the data and debugging components. With KNIME, reproducibility also plays an important role, e.g. by not removing old nodes. Therefore, old versions can still produce the same results later. Both offer a graphical user interface to abstract from the code.

The implementation of the functionality varies depending on the data type. The preparation process differs depending on whether you are working with structured or unstructured data. Other challenges come with streaming data. Data Civilizer 2.0 focuses on structured data. KNIME, on the other hand, also offers components for unstructured data.

Even though the main emphasis of this tools is more on analysis than on data preparation, approaches like these could provide a good starting point for the planned research.

\section{Test Data}
\label{sec:gouda}
At the beginning we have considered how to analyze and evaluate existing and new approaches. First, data that covers different application scenarios is needed. These should include many different types of errors that can occur in practice.

We encountered the problem that not enough suitable data sets were available for a detailed analysis, due to the following reasons:
\begin{itemize}
    \item \textbf{Non-public}: Some data sets are not publicly available. The reasons for this vary: some records are proprietary, for others no source is listed or the link provided is outdated.
    \item \textbf{Missing ground truth}: For some data sets, no ground truth is specified. Yet, this is a prerequisite for the evaluation of tools.
    \item \textbf{Error types}: Even when data sets are publicly available and ground truth is present, they usually contain only one or a few types of errors. For a comprehensive analysis of data cleaning tools, a significant range of different error types must be considered.
    \item \textbf{Error rate}: Real data normally contain rather a small number of errors. This limits the validity of the evaluation of data cleaning tools.
\end{itemize}

To overcome these problems, we have implemented a test data generator called \textit{GouDa - Generation of universal Data Sets}~\cite{gouda-paper}. Error types and rates are configurable with GouDa. Basis for this is the data generator implemented as part of the EvoBench project~\cite{Conrad2021}. We have extended the data generator to include the following properties:
\begin{itemize}
    \item \textbf{Error types}: A variety of different error types can be generated. The included errors are listed in~\autoref{tab:errortypes}. The classification is based on the definition of~\cite{oliveira2005} and has been extended to include further important error types, based on~\cite{Rahm2000, Kim2003, Mueller2005, Li2011}.
    \item \textbf{Error rate}: The rate is freely configurable.
    \item \textbf{Reproducible}: The generated data sets are reproducible insofar that the same type of error can be generated on the same attribute of the same tuple.
    \item \textbf{Scalable}: Various options for scaling are available.
    \item \textbf{Portable}: For a better portability, the data sets are generated in JSON format.
    \item \textbf{Realistic}: For data sets to be as realistic as possible, different lists with realistic names, words or addresses are included. The opportunity to include lists with possible attribute values is provided too.
    \item \textbf{Ground truth}: For each data set with errors, a data set with ground truth is provided. Additionally, the errors are logged in a text file.
\end{itemize}

\begin{table}[h]
  \caption{Error types supported by GouDa}
  \centering
  \label{tab:errortypes}
  %\begin{small}
  \begin{tabular}{ll}
    \toprule
    Level & Error Type\\
    \midrule
    \multirow{9}*{\shortstack[l]{An Attribute Value\\ of a Single Tuple}} & Missing value\\
     \cmidrule(r){2-2}
     & Syntax violation\\
     \cmidrule(r){2-2}
     & Interval violation\\
     \cmidrule(r){2-2}
     & Set violation\\
     \cmidrule(r){2-2}
     & Misspelled error\\
     \cmidrule(r){2-2}
     & Inadequate value to the attribute context\\
     \cmidrule(r){2-2}
     & Value items beyond the attribute context\\
     \cmidrule(r){2-2}
     & Meaningless Value\\
     \cmidrule(r){2-2}
     & Erroneous entry\\
     \midrule
     \multirow{4}*{\shortstack[l]{The Values\\ of a Single Attribute}} & Uniqueness value violation\\
     \cmidrule(r){2-2}
     & Synonyms existence\\
     \cmidrule(r){2-2}
     & Outlier*\\
     \cmidrule(r){2-2}
     & Missing Attribute\\
     \midrule
     \multirow{3}*{\shortstack[l]{The Attribute Values\\ of a Single Tuple}} & Semi-empty tuple\\
     \cmidrule(r){2-2}
     & Inconsistency among attribute values\\
     \cmidrule(r){2-2}
     & Irrelevant observation\\
     \midrule
     \multirow{4}*{\shortstack[l]{The Attribute Values\\ of Several Tuples}} & Redundancy about an entity\\
     \cmidrule(r){2-2}
     & Inconsistency about an entity\\
    \cmidrule(r){2-2}
     & Bias\\
    \cmidrule(r){2-2}
     & Noise\\
  \bottomrule
\end{tabular}
%\end{small}
\end{table}

\section{Conclusion and Outlook}
\label{sec:conclusion}
In this paper we have described current challenges in the context of data preparation pipelines. Derived from this, we have proposed a holistic data preparation tool. To accomplish the properties of such a tool, we have presented a number of aspects to be analyzed in our future research. These include data quality-related aspects, such as \emph{ensuring fairness} and a \emph{comprehensive integrated evaluation} as well as process-related aspects, such as \emph{abstraction from IT-knowledge}, \emph{minimizing human involvement}, \emph{reproducibility}, and \emph{scalibility}. We also want to investigate data-related aspects, such as \emph{semi-structured and unstructured data}, \emph{data changes} and \emph{streaming data}. As basis for our research we have introduced a data generator called \emph{GouDa -- Generation of universal Data Sets}. This permits the generation of erroneous data for comprehensive testing and evaluation.

We are currently using data sets generated by GouDa to investigate the following topics: First, we examine the current data preparation tools. We want to determine which tools are suitable for which data preparation steps and which error types. Secondly, since no single tool covers everything so far, we investigate the possible tool combinations for a most efficient data preparation pipeline. In context of this pipeline, we are currently focusing on data-related problems. Here we examine the different data formats. This concerns semi-structured and unstructured data and also streaming data. We deal with the question which special features these formats entail. Depending on this, we investigate which concepts for structured data can be transferred to these formats and where solutions must be adapted or newly developed. Following this, in future research we will address all topics described in Section~\ref{sec:research}.

\bibliographystyle{unsrt}  
\bibliography{template}  %%% Remove comment to use the external .bib file (using bibtex).

\begin{thebibliography}{10}

\bibitem{Stoyanovich2020}
Julia Stoyanovich, Bill Howe, and H.~V. Jagadish.
\newblock Responsible data management.
\newblock {\em Proc. VLDB Endow.}, page 3474–3488, 2020.

\bibitem{Klettke2021}
Meike Klettke and Uta St{\"{o}}rl.
\newblock {Four Generations in Data Engineering for Data Science: The Past,
  Presence and Future of a Field of Science}.
\newblock {\em Datenbank-Spektrum}, 2021.

\bibitem{Mahdavi2019}
Mohammad Mahdavi et~al.
\newblock {Towards Automated Data Cleaning Workflows}.
\newblock In {\em Proceedings of the Conference on "Lernen, Wissen, Daten,
  Analysen", Berlin, Germany, September 30 - October 2, 2019}, pages 10--19.
  CEUR-WS.org, 2019.

\bibitem{Li2021}
Peng Li et~al.
\newblock {CleanML: {A} Study for Evaluating the Impact of Data Cleaning on
  {ML} Classification Tasks}.
\newblock In {\em 37th {IEEE} International Conference on Data Engineering,
  {ICDE} 2021, Chania, Greece, April 19-22, 2021}, pages 13--24. {IEEE}, 2021.

\bibitem{katara}
Xu~Chu, John Morcos, Ihab~F. Ilyas, Mourad Ouzzani, Paolo Papotti, Nan Tang,
  and Yin Ye.
\newblock {KATARA:} {A} data cleaning system powered by knowledge bases and
  crowdsourcing.
\newblock In {\em Proceedings of the 2015 {ACM} {SIGMOD} International
  Conference on Management of Data, Melbourne, Victoria, Australia, May 31 -
  June 4, 2015}, pages 1247--1261. {ACM}, 2015.

\bibitem{dboost}
Yuxiao Zhang, Xiaorong Wang, Bingyang Li, Wei Chen, Tengjiao Wang, and Kai Lei.
\newblock Dboost: {A} fast algorithm for dbscan-based clustering on high
  dimensional data.
\newblock In {\em Advances in Knowledge Discovery and Data Mining - 20th
  Pacific-Asia Conference, {PAKDD} 2016, Auckland, New Zealand, April 19-22,
  2016, Proceedings, Part {II}}, pages 245--256. Springer, 2016.

\bibitem{Abedjan2016}
Ziawasch Abedjan et~al.
\newblock Detecting data errors: Where are we and what needs to be done?
\newblock {\em Proc. {VLDB} Endow.}, pages 993--1004, 2016.

\bibitem{Petrova_2020}
Dessislava Petrova-Antonova and Rumyana Tancheva.
\newblock Data cleaning: A case study with openrefine and trifacta wrangler.
\newblock In {\em Quality of Information and Communications Technology}, pages
  32--40, Cham, 2020. Springer International Publishing.

\bibitem{holistic-paper}
Valerie Restat, Meike Klettke, and Uta St{\"{o}}rl.
\newblock {Towards a Holistic Data Preparation Tool}.
\newblock In {\em DataPlat, 25th International Conference on Extending Database
  Technology (EDBT), 2022, Edinburgh, UK}, 2022.

\bibitem{Ridzuan2019}
Fakhitah Ridzuan and Wan Mohd~Nazmee {Wan Zainon}.
\newblock {A Review on Data Cleansing Methods for Big Data}.
\newblock {\em Procedia Computer Science}, pages 731--738, 2019.
\newblock The Fifth Information Systems International Conference, 23-24 July
  2019, Surabaya, Indonesia.

\bibitem{Krishnan2016}
Sanjay Krishnan et~al.
\newblock Towards reliable interactive data cleaning: a user survey and
  recommendations.
\newblock In {\em Proceedings of the Workshop on Human-In-the-Loop Data
  Analytics, HILDA@SIGMOD 2016, San Francisco, CA, USA, June 26 - July 01,
  2016}, page~9. {ACM}, 2016.

\bibitem{Hameed2020}
Mazhar Hameed and Felix Naumann.
\newblock {Data Preparation: {A} Survey of Commercial Tools}.
\newblock {\em {SIGMOD} Rec.}, pages 18--29, 2020.

\bibitem{Pawlik2019}
Mateusz Pawlik et~al.
\newblock {A Link is not Enough - Reproducibility of Data}.
\newblock {\em Datenbank-Spektrum}, pages 107--115, 2019.

\bibitem{Rupprecht2020}
Lukas Rupprecht et~al.
\newblock {Improving Reproducibility of Data Science Pipelines through
  Transparent Provenance Capture}.
\newblock {\em Proc. {VLDB} Endow.}, pages 3354--3368, 2020.

\bibitem{Cai2015}
Li~Cai and Yangyong Zhu.
\newblock {The Challenges of Data Quality and Data Quality Assessment in the
  Big Data Era}.
\newblock {\em Data Sci. J.}, page~2, 2015.

\bibitem{Sidi2012}
Fatimah Sidi et~al.
\newblock Data quality: {A} survey of data quality dimensions.
\newblock In {\em 2012 International Conference on Information Retrieval {\&}
  Knowledge Management, Kuala Lumpur, Malaysia, March 13-15, 2012}, pages
  300--304. {IEEE}, 2012.

\bibitem{Pitoura2020}
Evaggelia Pitoura.
\newblock {Social-minded Measures of Data Quality: Fairness, Diversity, and
  Lack of Bias}.
\newblock {\em {ACM} J. Data Inf. Qual.}, pages 12:1--12:8, 2020.

\bibitem{Klettke2021_2}
Meike Klettke, Adrian Lutsch, and Uta St{\"{o}}rl.
\newblock Kurz erkl{\"{a}}rt: Measuring data changes in data engineering and
  their impact on explainability and algorithm fairness.
\newblock {\em Datenbank-Spektrum}, pages 245--249, 2021.

\bibitem{Schelter2020}
Sebastian Schelter and Julia Stoyanovich.
\newblock Taming technical bias in machine learning pipelines.
\newblock {\em IEEE Data Engineering Bulletin (Special Issue on
  Interdisciplinary Perspectives on Fairness and Artificial Intelligence
  Systems)}, pages 39--50, 2020.

\bibitem{Mehrabi2021}
Ninareh Mehrabi et~al.
\newblock {A Survey on Bias and Fairness in Machine Learning}.
\newblock {\em {ACM} Comput. Surv.}, pages 115:1--115:35, 2021.

\bibitem{Seifert2019}
Christin Seifert, Stefanie Scherzinger, and Lena Wiese.
\newblock {Towards Generating Consumer Labels for Machine Learning Models}.
\newblock In {\em 2019 {IEEE} First International Conference on Cognitive
  Machine Intelligence (CogMI), Los Angeles, CA, USA, December 12-14, 2019},
  pages 173--179. {IEEE}, 2019.

\bibitem{Christ2016}
James Crist.
\newblock Dask {\&} numba: Simple libraries for optimizing scientific python
  code.
\newblock In {\em 2016 {IEEE} International Conference on Big Data {(IEEE}
  BigData 2016), Washington DC, USA, December 5-8, 2016}, pages 2342--2343.
  {IEEE} Computer Society, 2016.

\bibitem{Petersohn2021}
Devin Petersohn et~al.
\newblock Flexible rule-based decomposition and metadata independence in modin:
  {A} parallel dataframe system.
\newblock {\em Proc. {VLDB} Endow.}, pages 739--751, 2021.

\bibitem{Chu2016}
Xu~Chu et~al.
\newblock {Data Cleaning: Overview and Emerging Challenges}.
\newblock In {\em Proceedings of the 2016 International Conference on
  Management of Data, {SIGMOD} Conference 2016, San Francisco, CA, USA, June 26
  - July 01, 2016}, pages 2201--2206. {ACM}, 2016.

\bibitem{Ditzler2013}
Gregory Ditzler and Robi Polikar.
\newblock Incremental learning of concept drift from streaming imbalanced data.
\newblock {\em {IEEE} Trans. Knowl. Data Eng.}, pages 2283--2301, 2013.

\bibitem{Volkovs2014}
Maksims Volkovs et~al.
\newblock Continuous data cleaning.
\newblock In {\em {IEEE} 30th International Conference on Data Engineering,
  Chicago, {ICDE} 2014, IL, USA, March 31 - April 4, 2014}, pages 244--255.
  {IEEE} Computer Society, 2014.

\bibitem{Rezig2019}
El~Kindi Rezig et~al.
\newblock Data civilizer 2.0: {A} holistic framework for data preparation and
  analytics.
\newblock {\em Proc. {VLDB} Endow.}, pages 1954--1957, 2019.

\bibitem{Fillbrunn2017}
Alexander Fillbrunn et~al.
\newblock Knime for reproducible cross-domain analysis of life science data.
\newblock {\em Journal of Biotechnology}, pages 149--156, 2017.
\newblock Bioinformatics Solutions for Big Data Analysis in Life Sciences
  presented by the German Network for Bioinformatics Infrastructure.

\bibitem{gouda-paper}
Valerie Restat, Gerrit Boerner, Andr{\'{e}} Conrad, and Uta St{\"{o}}rl.
\newblock Gouda - generation of universal data sets: improving analysis and
  evaluation of data preparation pipelines.
\newblock In {\em {DEEM} '22: Proceedings of the Sixth Workshop on Data
  Management for End-To-End Machine Learning Philadelphia, PA, USA, 12 June
  2022}, pages 2:1--2:6. {ACM}, 2022.

\bibitem{Conrad2021}
Andr{\'{e}} Conrad et~al.
\newblock Evobench: Benchmarking schema evolution in nosql.
\newblock In {\em Performance Evaluation and Benchmarking - 13th {TPC}
  Technology Conference, {TPCTC} 2021, Copenhagen, Denmark, August 20, 2021,
  Revised Selected Papers}, pages 33--49. Springer, 2021.

\bibitem{oliveira2005}
Paulo Oliveira et~al.
\newblock A taxonomy of data quality problems.
\newblock In {\em 2nd Int. Workshop on Data and Information Quality}, pages
  219--233, 2005.

\bibitem{Rahm2000}
Erhard Rahm and Hong~Hai Do.
\newblock Data cleaning: Problems and current approaches.
\newblock {\em {IEEE} Data Eng. Bull.}, pages 3--13, 2000.

\bibitem{Kim2003}
Won~Y. Kim et~al.
\newblock A taxonomy of dirty data.
\newblock {\em Data Min. Knowl. Discov.}, pages 81--99, 2003.

\bibitem{Mueller2005}
Heiko M{\"u}ller and Johann~Christoph Freytag.
\newblock Problems, methods, and challenges in comprehensive data cleansing.
\newblock Technical report hub-ib-164, Humboldt University, 2003.

\bibitem{Li2011}
Lin Li, Taoxin Peng, and Jessie Kennedy.
\newblock A rule based taxonomy of dirty data.
\newblock {\em GSTF Journal on Computing (JoC)}, 2011.

\end{thebibliography}

\end{document}